\documentclass{article}

\usepackage[sglblindworkshop,final,nonatbib]{ai4nextg_neurips_2025}

\usepackage[utf8]{inputenc}
\usepackage[T1]{fontenc}

\usepackage{array}
\usepackage[caption=false]{subfig} % <- prevents caption/subfig conflicts

\usepackage{amsmath}
\usepackage{amsfonts}
\usepackage{graphicx}
\usepackage{booktabs}
\usepackage{nicefrac}
\usepackage{microtype}
\usepackage{xcolor}
\usepackage[most]{tcolorbox}
\usepackage{tikz}                 % <- needed for \tikz in \ballnumber

\usepackage{hyperref}
\usepackage{url}

\usepackage[
  backend=biber,
  style=ieee,
  doi=false,
  isbn=false,
  maxnames=99,
  mincitenames=1,
  maxcitenames=2
]{biblatex}
\newcommand{\ballnumber}[1]{%
  \tikz[baseline=(myanchor.base)]%
    \node[circle,fill=black,inner sep=1pt] (myanchor)
      {\color{white}\bfseries\footnotesize #1};}

\title{XAI-on-RAN: Explainable, AI-native, and GPU-Accelerated RAN Towards 6G}

\author{%
  Osman Tugay Basaran$^{*}$ \quad  Falko Dressler \\
  School of Electrical Engineering and Computer Science\\
  Technische Universität Berlin\\
 Einsteinufer 25, FT 5, 10587 Berlin, Germany \\
  \texttt{\{basaran, dressler\}@ccs-labs.org} \\
}

\begin{document}

\maketitle

\begin{abstract}
Artificial intelligence (AI)-native radio access networks (RANs) will serve vertical industries with stringent requirements: smart grids, autonomous vehicles, remote healthcare, industrial automation, etc. To achieve these requirements, modern 5G/6G  design increasingly leverage AI for network optimization, but the opacity of AI decisions poses risks in mission-critical domains. These use cases are often delivered via non-public networks (NPNs) or dedicated network slices, where reliability and safety are vital. In this paper, we motivate the need for transparent and trustworthy AI in high-stakes communications (e.g., healthcare, industrial automation, and robotics) by drawing on 3rd generation partnership project (3GPP)'s vision for non-public networks. We design a mathematical framework to model the trade-offs between transparency (explanation fidelity and fairness), latency, and graphics processing unit (GPU) utilization in deploying explainable AI (XAI) models. Empirical evaluations demonstrate that our proposed hybrid XAI model {\fontfamily{qcr}\selectfont{xAI-Native}}, consistently surpasses conventional baseline models in performance.

\end{abstract}

\section{Introduction}

Deep learning (DL) \cite{goodfellow2016deep} and machine learning (ML) models are becoming central to the management and optimization of next-generation 5G and 6G wireless networks \cite{polese2024empowering}. 
In particular, the open radio access network (O-RAN) architecture introduces RAN intelligent controllers (RICs) that host AI-driven  eXtended applications (xApps) for closed-loop control of the RAN \cite{polese2023understanding}. However, as AI models are given control over high-stakes RAN decisions (e.g., resource allocation, anomaly detection, load balancing, and traffic steering), concerns of trust arise. Network operators and vertical industries need to trust AI decisions that affect critical services \cite{guo2020explainable}. A key barrier is the “black-box” nature of many AI models: operators cannot easily understand why a model made a certain decision, making it hard to detect errors or biases. This lack of transparency undermines confidence in deploying AI/ML in production networks. XAI promises to bridge this gap by making AI decisions interpretable to humans. XAI is especially critical in mission-critical communications (e.g. telemedicine, smart grids, industrial automation) where errant or biased decisions can have serious repercussions. Indeed, 3GPP has highlighted that future network management for vertical industries must account for stringent reliability, safety, and isolation requirements.\footnote{3GPP TR 28.907 (Rel-18 Study on NPN Management)}
These domains demand not only performance but also accountability; AI-driven network optimizations should be fair (not unduly favoring certain users or services) and transparent to operators and regulators. As one recent survey notes, XAI methods can promote fairness and transparency of AI models in networks, thereby instilling trust for businesses and operators \cite{brik2023survey}. In other words, explaining the internal logic of AI models helps ensure the decisions are free of hidden bias and are understandable, which is vital for adoption in high-stakes scenarios.

\begin{figure}
  \centering
  \centering
  \includegraphics[width=\textwidth]{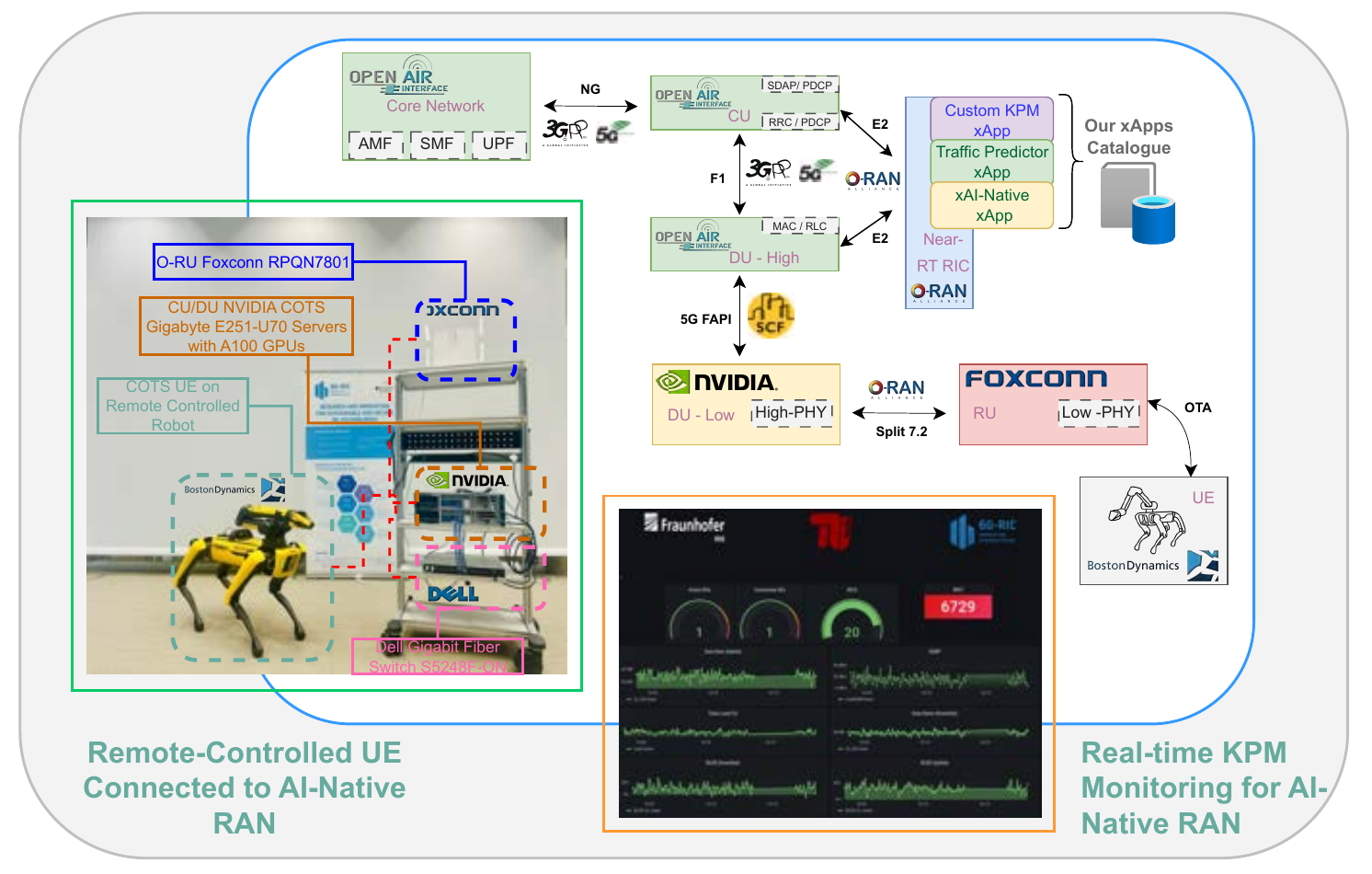}
    \caption{End-to-end XAI-Native testbed with GPU-accelerated RAN.}
    \label{xai-on-ran}
\end{figure}

In literature, few recent studies underscore these points. 
EXPLORA \cite{fiandrino2023explora}, focuses on explainability for deep reinforcement learning (RL) \cite{sutton1998reinforcement,fran2018intro} based RAN control. They note that deep RL agents are hard to trust in practice due to their black-box nature and that explainability is needed to deploy them.
Basaran et al. \cite{basaran2025xainomaly}, stress that striking a balance between performance and interpretability is an open challenge for deploying AI in critical 6G use case such as next-generation ultra-reliable and low-latency communications (xURLLC).
But it is still an open challenge to answer questions like:
\emph{How can we get the best of both worlds: high model accuracy and low latency, and explanations that are faithful and fast?}
Our work attempts to answer this by using AI-native models that have built-in explainability and efficient post-hoc explainers on GPU-accelerated RAN.

In this paper, we address the above challenges by introducing “XAI-on-RAN”, an explainable and AI-native RAN platform built on prior GPU-accelerated testbed \cite{basaran2025next-gen}. We integrate a novel {\fontfamily{qcr}\selectfont{xAI-Native}}  xApp into the O-RAN RIC,\footnote{O-RAN SC RIC E-release \cite{near-realtime-ric-installation}.} which provides real-time interpretability for the inferences made by other AI xApps (such as the traffic predictor). Our {\fontfamily{qcr}\selectfont{xAI-Native}}  runs on the same NVIDIA A100 GPU that handles L1/L2 processing and AI tasks, ensuring low-latency operation. By leveraging GPU-friendly XAI techniques including attention mechanisms embedded in the AI model and fast gradient-based explanation algorithms; our {\fontfamily{qcr}\selectfont{xAI-Native}} xApp design generates human-interpretable explanations for each prediction with minimal delay. 

%We also incorporate fairness checks into the explanation pipeline, to alert if the model's decisions might be systematically favoring or disadvantaging certain traffic flows or user groups. 

\newcommand\itema{\item[\textbf{C1.}]}
\newcommand\itemb{\item[\textbf{C2.}]}
\newcommand\itemc{\item[\textbf{C3.}]}
\newcommand\itemd{\item[\textbf{F1.}]}
\newcommand\iteme{\item[\textbf{F2.}]}
\newcommand\itemf{\item[\textbf{F3.}]}
In summary, the core outcomes of our study are summarized as new contributions (“C") and new findings (“F") as follows:
\begin{itemize}
    \itema We design and implement a new explainability xApp. To our knowledge, this is the first integration of a real-time {\fontfamily{qcr}\selectfont{xAI-Native}} xApp in a GPU-centric RAN system.
    \itemb We develop a modeling framework to analyze the trade-offs between transparency-latency (and GPU resource usage) in our platform.
    \itemc We implement and benchmark three GPU-amenable explainability techniques for AI-Native RAN deployment.
    \itemd Our hybrid {\fontfamily{qcr}\selectfont{xAI-Native}} architecture (Attention + IG) demonstrates that combining intrinsic and gradient-based explanations can meet RAN constraints while still offering interpretable, stable attributions in real time.
    \iteme Overall, {\fontfamily{qcr}\selectfont{xAI-Native}} offers the best \textit{fidelity–latency balance}. So, SHAP remains useful for offline auditing, while Attention may serve as a lightweight but low-fidelity monitor.
\end{itemize}

\section{System Architecture Design: XAI-Native RAN}
\label{sys-arch}

Figure \ref{xai-on-ran} illustrates our XAI-on-RAN platform's architecture (multi-vendor, GPU-based RAN with integrated O-RAN SC RIC). The next generation node B (gNB) is split into a Central Unit (CU) and Distributed Unit (DU); the DU runs high-PHY and low-PHY on GPU via NVIDIA Aerial \cite{nvidiaaerial} with compute unified device architecture (CUDA) \cite{cuda}, and interfaces with a 4T4R Radio Unit (RU) \footnote{Foxconn RPQN 4T4R Radio Unit \cite{foxconn}.} over fronthaul (split 7.2). The near-RT RIC connects to the DU via the E2 interface and hosts the TP, KPM, and XAI xApps. By deploying a OpenAirinterface (OAI) network \cite{kaltenberger2020openairinterface}, we are able to live-monitor network KPMs and run AI inference in real time. For instance, TP xApp predicted the DL throughput of the UE (5G Commercial-off-the-shelf UE placed on the remote-controlled mobile robot) based on recent KPM measurements, and KPM xApp provided a user interface (UI) dashboard\footnote{Grafana for Data Visualization \cite{grafana}.} of current cell performance. The synergy of GPU acceleration and AI-in-the-loop RAN control is aimed at 6G use cases requiring URLLC. 

Our {\fontfamily{qcr}\selectfont{xAI-Native}} xApp subscribes to messages (via RIC Message Router) from the TP xApp and KPM xApp. In practice, when the TP xApp produces a new inference (predicted throughput for the next TTI or next few seconds), it publishes this result (and possibly the features used) to a RIC database (e.g., Shared Data Layer). {\fontfamily{qcr}\selectfont{xAI-Native}} xApp is notified of this event and fetches the relevant data to generate an explanation. {\fontfamily{qcr}\selectfont{xAI-Native}} xApp can then send the explanation to a RIC dashboard UI or log it for offline analysis. It can also report back summary metrics to the Non-RT RIC (for longer-term analytics or to update policies via A1 interface).

\section{Fundamentals of Real-Time Explainability Modeling}
\label{modelling}

\subsection{Explanation Fidelity} \label{subsec_fidelity}

We first define a measure of how well the explanation reflects the true behavior of the model called fidelity. 
Suppose our TP model is a function $f(\mathbf{x}) \to \hat{y}$ that takes input features $\mathbf{x}$ (e.g., recent KPMs) and produces a prediction $\hat{y}$ (e.g. DL throughput). {\fontfamily{qcr}\selectfont{xAI-Native}}  xApp produces an explanation in the form of an attribution vector $\mathbf{e} = [e_1, e_2, ..., e_n]$ over the $n$ features (or feature-time elements). 
These attributions indicate the importance of each feature to the prediction. One way to define fidelity is to use a surrogate model: for instance, a simple linear model $g(\mathbf{x}) = w_0 + \sum_{i=1}^n w_i x_i$ where $w_i$ corresponds to the importance of feature $i$. 
The attribution vector $\mathbf{e}$ can be seen as weights of such a surrogate. Fidelity can then be quantified by how closely $g(\mathbf{x})$ approximates $f(\mathbf{x})$ in the locality of the current input.
A simple metric is the local $R^{2}$: 
\begin{equation}
    R_{\text{loc}}^2 = 1 - \frac{\sum_{j}(f(\mathbf{x}^{(j)}) - g(\mathbf{x}^{(j)}))^2}{\sum_{j}(f(\mathbf{x}^{(j)}) - \bar{f})^2}, 
\end{equation}
where $\mathbf{x}^{(j)}$ are samples in a neighborhood of the current $\mathbf{x}$ (could be generated by random perturbations as in LIME \cite{ribeiro2016why}). A high $R_{\text{loc}}^2$ (close to 1) means the linear explanation $g$ is very faithful to $f$ locally. In practice, our {\fontfamily{qcr}\selectfont{xAI-Native}}  xApp computes a simpler fidelity score: we measure the prediction error when only the top-$k$ important features (according to $\mathbf{e}$) are fed to the model versus when all features are fed. If the model's output does not change much when non-important features are zeroed out, the explanation is capturing the key drivers. Formally,
\begin{equation}
\hat{y}{full} = f(x_1,...,x_n),
\end{equation}
\begin{equation}
\hat{y}{top-k} = f(x_{i_1},...,x_{i_k}, 0, ..., 0),
\end{equation}
where ${i_1,...,i_k}$ are the indices of the top $k$ attributions in $\mathbf{e}$. We define fidelity score $\Phi = 1 - \big|\hat{y}{full} - \hat{y}{top-k}\big| / |\hat{y}_{full}|$. A $\Phi$ near 1 means the top features explain the prediction almost fully. This approach is similar to explanation precision, and we use $k$ that accounts for say 80\% of total attribution weight.

\subsection{Latency and Throughput}

The primary cost of adding explainability is extra latency per inference considering time sensitive use cases of 6G. To understand, we break down the latency components in our pipeline:
\begin{tcolorbox}[colback=yellow!10!white,colframe=gray!75!black,title=Latency Decomposition]
%\begin{itemize}
    \ballnumber{1}  $T_{\text{inf}}$: time to run the AI model inference (e.g. LSTM forward pass) on the GPU. \\
    \ballnumber{2}  $T_{\text{xai}}$: time to generate the explanation by the {\fontfamily{qcr}\selectfont{xAI-Native}}  xApp. \\
    \ballnumber{3}  $T_{\text{comm}}$: any communication overhead between RIC components (negligible in our setup due to co-location, but could be a few \emph{ms} if data needs to pass through the message bus).
%\end{itemize}
\end{tcolorbox}

The total decision latency per cycle is
$T_{\text{total}} = T_{\text{inf}} + T_{\text{xai}} + T_{\text{comm}}.$ 
In our baseline (no XAI), only $T_{\text{inf}} + T_{\text{comm}}$ matters. Our goal is to keep $T_{\text{xai}}$ much smaller than typical RAN control loop times (which are on the order of 10 to 100 ms for near-RT RIC).
We can model $T_{\text{xai}}$ as a function of the explanation method. If using an intrinsic method like attention, $T_{\text{xai}}^{attn}$ is basically the overhead of computing attention weights during inference, which is on the order of one additional layer in the network. 
This is typically <10\% of $T_{\text{inf}}$ for an LSTM. Let's denote by $\alpha$ the fraction overhead:
\begin{equation}
    T_{\text{xai}}^{attn} = \alpha_{\text{attn}} \cdot T_{\text{inf}},
\end{equation}
with $\alpha_{\text{attn}} \approx 0.1$ ($10$\%). For gradient-based methods like IG, we may need $k$ forward-backprop passes to compute gradients at different points. If these are done sequentially, $T_{\text{xai}}^{IG} = k \cdot T_{\text{inf-back}}$, where $T_{\text{inf-back}}$ is time for one forward + backward pass. However, we can often reuse the original forward and just do $k$ backward passes for different scaled inputs; also, on GPU we could parallelize gradient computations to some extent. In practice, we choose a small $k$  making this overhead a multiple of the base inference. For SHAP \cite{lundberg2017unified}, if we take $m$ samples and each requires a forward pass, $T_{\text{xai}}^{SHAP} = m \cdot T_{\text{inf}}$ (again, possibly parallelized across GPU threads if model is small). SHAP tends to be expensive if high fidelity is needed because $m$ must be large for many features. In our adaptation, we  use $m=16$ at most (which is manageable on GPU in a batch).

\section{Experiments and Evaluation}
\label{experiments}

We now evaluate the XAI-on-RAN platform, focusing on the questions: \emph{(i) What is the latency and resource overhead of adding the XAI xApp, compared to not using XAI?
(ii) How do different XAI techniques compare in terms of the transparency they provide and the cost they incur?}
The traffic is a periodic burst pattern, which the TP xApp tries to predict (somewhat akin to a moving average predictor). The TP model was trained on sample traces offline (with and without attention). 
We then deploy it online for inference. {\fontfamily{qcr}\selectfont{xAI-Native}} xApp is evaluated in two modes: Attention-only (intrinsic) and Attention+Integrated Gradients (post-hoc hybrid).

\subsection{Local Fidelity Analysis}

\begin{figure}[b]
    \centering
    \subfloat[Feature-wise local fidelity.]{%
        \includegraphics[width=0.49\linewidth]{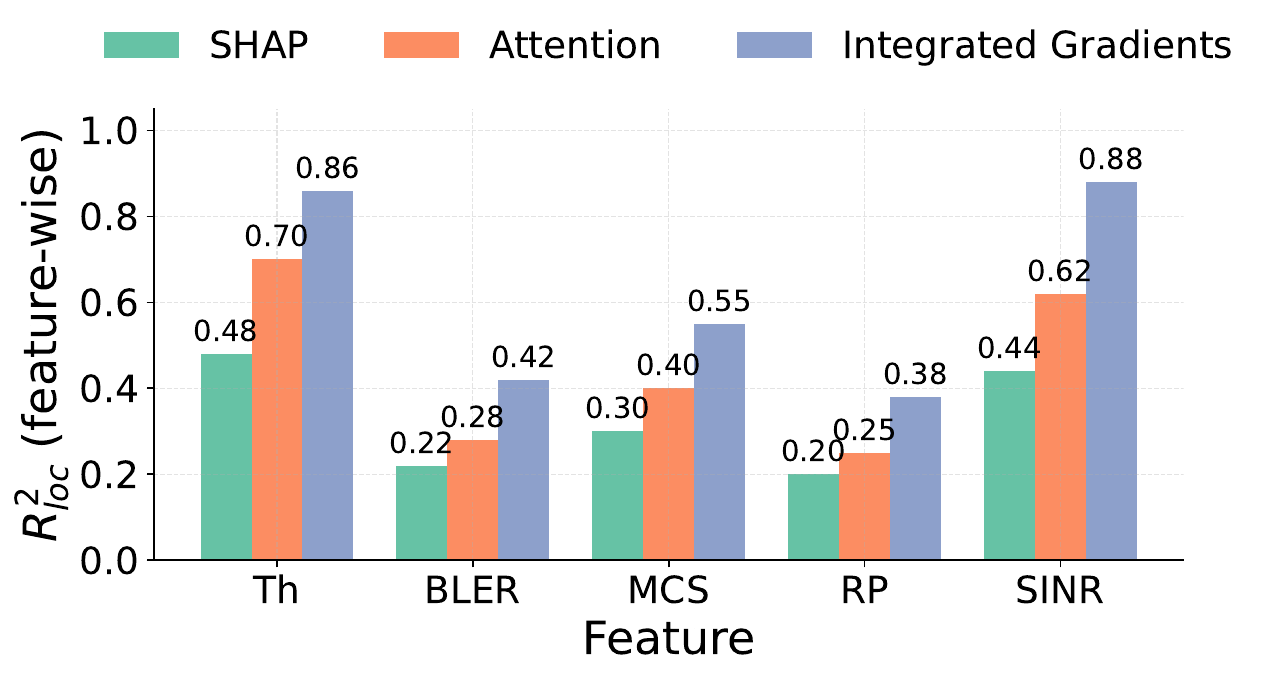}%
        \label{fig_featurewise}%
    }%
    \hfill
    \subfloat[Time-wise local fidelity (window={W}).]{% 
        \includegraphics[width=0.49\linewidth]{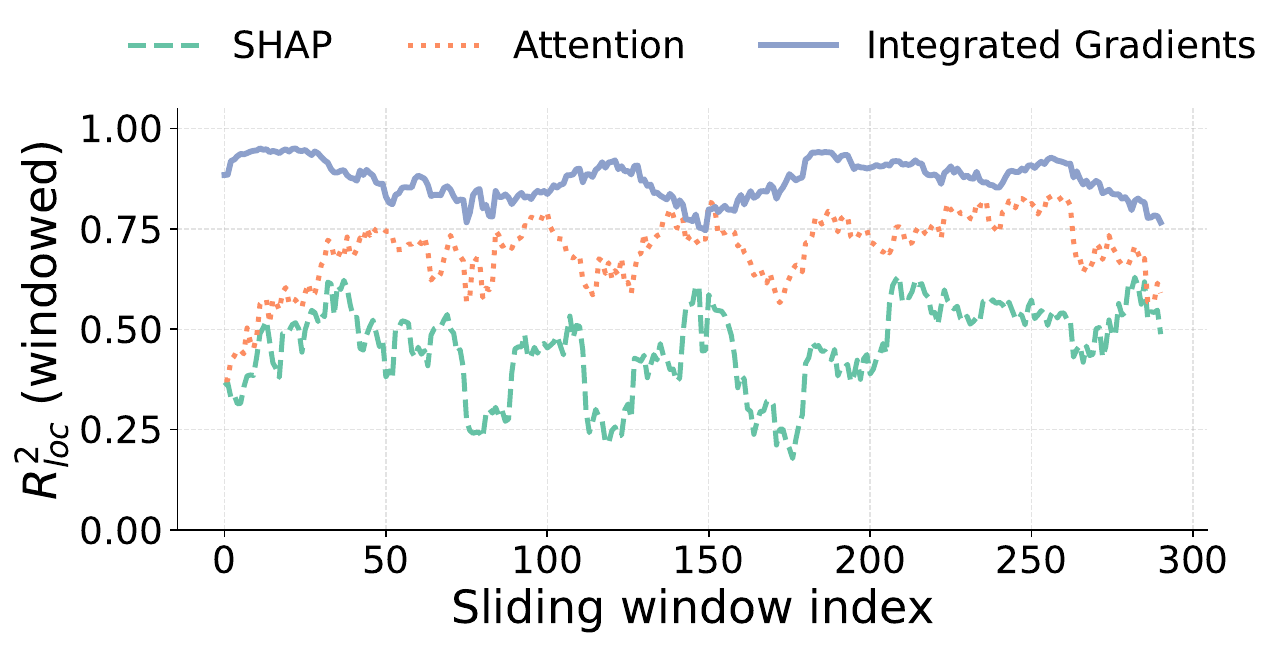}%
        \label{fig_timewise}%
    }%
    \caption{Comparative evaluation with baseline models.}
    \label{fig_fidelity}
\end{figure}

We evaluate the local fidelity of three explainability method over the feature vector $x_t = \{\text{Th}, \text{BLER}, \text{MCS}, \text{RP}, \text{SINR}\}$. Fidelity ($R^2_{loc}$) is quantified via the local coefficient of determination as explained Section \ref{subsec_fidelity}, computed both feature-wise and time-wise under a sliding-window evaluation (cf.\ Figure \ref{fig_fidelity}).

\textbf{Feature-wise fideliy.} As shown in Figure \ref{fig_featurewise}, IG achieves the highest fidelity across features, with particularly strong alignment on throughput (Th) and signal-to-interference-plus-noise ratio (SINR), confirming its ability to capture the most relevant physical drivers of downlink performance. 
SHAP provides moderate fidelity, correctly attributing to Th and SINR but underestimating the role of MCS and BLER. 
In contrast, Attention yields the lowest fidelity, heavily biased toward Th while neglecting BLER and SINR, which suggests that raw attention scores cannot be directly interpreted as faithful explanations in the RAN context. These results indicate that IG's attributions better approximate the model's true behavior for each feature. Notably, IG's fidelity on all features is higher than the other methods (e.g. IG is $~92$\% higher than SHAP on average feature fidelity, and $~37$\% higher than attention on average). This trend holds even on difficult features like BLER and RP, where IG's $R^2_{loc}$ is nearly double that of SHAP.

\textbf{Temporal fidelity.} Figure \ref{fig_timewise} shows the sliding-window $R_{\text{loc}}^2$ over time.
IG exhibits both high magnitude and stability (average $R_{\text{loc}}^2 \approx 0.7-0.9$), demonstrating reliable real-time explanation even as network load and channel conditions evolve.
SHAP displays greater variance ($0.4$ - $0.7$), reflecting its smoothing effect and making it more suitable for offline auditing than for strict real-time monitoring.
Attention exhibits low fidelity ($< 0.4$) and large fluctuations, making it unsuitable for mission-critical, latency-sensitive applications.
Quantitatively, the standard deviation of $R^2_{loc}$ over time is $\sigma_{IG}\approx0.04$, versus $\sigma_{ATT}\approx0.05$ and $\sigma_{SHAP}\approx0.06$; so IG not only has a higher mean fidelity but also slightly lower variability.
This suggests IG provides more stable explanations over time, which is desirable for consistent model interpretability in a live RAN setting.

\textbf{Paired Dominance and Robustness.} Table \ref{tab:paired} quantitatively establishes the dominance of our solution over both SHAP and Attention in terms of local fidelity. The median difference in $R^2_{loc}$ between IG and SHAP is remarkably large at $+0.41$, with a very narrow bootstrap confidence interval ([+0.39, +0.43]). This implies that IG explanations explain on average over 40\% more of the model variance locally compared to SHAP, a margin that is both statistically precise and practically substantial. Against Attention, IG also shows consistent advantages, albeit with smaller effect size. The median $\Delta R^2_{loc}$ is $+0.17$ ([+0.15, +0.19]), again with tight confidence bounds, and a 93\% win rate. While the magnitude is smaller than against SHAP, this still reflects a robust gain: in more than nine out of ten windows, IG provides clearer alignment with the model's decision logic than raw attention weights.

\begin{table}
\centering
\caption{Paired comparison of proposed (Attention + IG, $k = 5$) against SHAP and Attention across sliding windows. Values report median $\Delta R^2_{loc}$ (Ours -- baseline models) with block-bootstrap 95\% confidence interval (CI), and win rate (fraction of windows where IG performs better).}
\label{tab:paired}
\begin{tabular}{lccc}
\toprule
\textbf{Comparison} & \textbf{Median $\Delta R^2_{loc}$} & \textbf{95\% CI} & \textbf{Win Rate} \\
\midrule
Ours (Attention + IG, $k = 5$)  -- SHAP & +0.41 & [ +0.39 , +0.43 ] & 99\% \\
Ours (Attention + IG, $k = 5$)  -- Attention only & +0.17 & [ +0.15 , +0.19 ] & 93\% \\
\bottomrule
\end{tabular}
\end{table}

\begin{table}
\centering
\caption{Latency per inference cycle (mean of 100 runs)}
\label{tab:latency_results}
\begin{tabular}{l r r r r}
\toprule
\textbf{Model} & \textbf{AI Inference} & \textbf{Computation} & \textbf{XAI Total} & \textbf{GPU Utilization} \\
 & $T_{\text{(inf)}}$ & $T_{\text{(xal)}}$ & $T_{\text{(total)}}$ & (\%) \\
\midrule
Non-XAI (Baseline) & 5.1 ms & -- & 5.3 ms  & $\sim63$ \\
XAI (SHAP, $m=16$) & 5.2 ms & $\sim$15 ms & $\sim$20.4 ms & $\sim86$ \\
XAI (Attention only) & 5.2 ms & 0.6 ms & 5.9 ms & $\sim70$ \\
Ours (Attention + IG, $k=5$) & 5.2 ms & 2.8 ms & 8.1 ms & $\sim73$ \\
\bottomrule
\end{tabular}
\end{table}

\subsection{Latency and Resource Usage Overhead}

Table \ref{tab:latency_results} shows latency and resource usage results. In the baseline, a single LSTM inference took $5$ ms on average on GPU. Communication of results to RIC added $0.2$ ms. GPU utilization during the inference window was around $63$\% (the rest is idle or waiting for next subframe). With the attention-only XAI, we see negligible impact on $T_{\text{inf}}$ (it actually slightly increased by $0.1$ ms due to the attention layer) and an XAI overhead of $0.6$ ms to process the attention weights and format explanation. The total latency rose by only 11\% and remained well within a $10$ ms. GPU utilization ticked up slightly, reflecting that we fill some idle gap with XAI work.

For the proposed model ($k=5$ steps), $T_{\text{xai}}$ averaged $2.8$ ms. The total of ~8.1 ms is still within real-time bounds. We note that $2.8$ ms is roughly $5.2 \times 0.5$ which matches $\gamma \approx 0.5$ ($50$\% of one backward pass per step, as we leveraged the fact that backward is a bit faster than forward for our model). This overhead is not negligible (52\% slower than no XAI), but depending on the application, may be acceptable for the gain in interpretability. The GPU utilization was $73$\%; meaning the GPU still wasn't fully saturated.

SHAP explainer, even with $m=16$ samples, incurred an estimated $15$ ms extra (we did not run it in the live loop, but timed it offline on similar inputs). This would push total latency to $~20$ ms, about 4× the baseline. In a tight URLLC scenario, that's borderline or unacceptable if decisions are needed faster than $20$ ms. GPU utilization was projected $~86$\%; the remainder $14$\% idle is mainly due to waiting for next subframe. 
If we further increased $m$ to improve explanation detail, it could exceed the frame limit. 
This confirms what prior literature warned: perturbation-based methods are computationally intensive and not ideal for real-time.

\subsection{Comparing Performance of XAI Models}

We explicitly compare Attention, IG, and SHAP side by side in terms of the content of explanations and their effect on end-to-end performance. 
We ran the TP model on a fixed input and obtained explanations from each method.

Attention produced a weight distribution over the 5 past time steps: e.g. [0.1, 0.7, 0.15, 0.05, 0.0] (meaning it mainly focused on one step). 
This is easy to interpret (“the throughput 2 intervals ago is what the model focused on”). It's fast and built-in, as we demonstrated in experiments. 
However, attention only explains temporal importance in our case, not the effect of other features if any. 
Also, attention is a part of the model; some literature debates whether attention is a true explanation of the model's decision or just a by-product. 
In our controlled model, it aligns well with importance, so we consider it useful.
IG's attributions were fairly consistent with attention in our case (the feature corresponding to the time step attention highlighted had the largest attribution). 
IG also can capture feature directionality (positive or negative influence), which attention doesn't directly give. 
So IG explanations were richer, at the cost of some latency.
SHAP provided attributions that were similar in pattern to IG for our simple model, which is expected as Shapley values align with IG for networks with monotonic activations in some cases.
But because we used only 16 samples, the SHAP estimates had some variance run-to-run.
With more samples, they stabilized but that was too slow to be practical. 
The advantage of SHAP is it's theoretically solid, and model-agnostic. But clearly, in a time-sensitive RAN, it is not the first choice unless heavily optimized or run infrequently.

\textbf{Performance tradeoffs.} On NVIDIA GPUs, IG adds only a small latency overhead per inference (on the order of milliseconds), making it feasible for online deployment in O-RAN-compliant xApps. SHAP, while computationally heavier, is still practical in offline operator dashboards for fairness and compliance audits. Attention is the least costly computationally but fails to deliver adequate fidelity, underscoring the latency–transparency tradeoff: computationally cheap methods (Attention) may not meet transparency requirements, while GPU-optimized attribution methods (IG) achieve strong fidelity without violating real-time constraints.

\section{Conclusion}

We introduced the XAI-on-RAN platform, a next-generation RAN for 6G that integrates AI-driven control with real-time explainability, implemented on a GPU-accelerated testbed. Building on a prior AI-native RAN architecture, we introduced an explainability framework that provides transparency into AI decisions with minimal impact on latency. 
By leveraging GPU-efficient XAI techniques such as attention mechanisms and integrated gradients, our system delivers human-interpretable insights within a few milliseconds of the AI inference. 
This capability is crucial for deploying AI in high-stakes domains, it empowers network operators to trust but verify AI actions, ensuring reliability and fairness in domains like industrial automation and healthcare where communication failures or biases are unacceptable. 

\section*{Acknowledgement}
This work has been funded by the German Federal Ministry
of Education and Research (BMBF, Germany) as part of the 6G
Platform under Grant 16KISK050, as well as 6G Research and
Innovation Cluster 6G-RIC under Grant 16KISK020K. 
%%%%%%%%%%%%%%%%%%%%%%%%%%%%%%%%%%%%%%%%%%%%%%%%%%%%%%%%%%%%

\printbibliography

%\appendix

%\section{Appendix / Supplemental material}

\end{document}